\documentclass[twocolumn,superscriptaddress,showpacs,preprintnumbers,amsmath,amssymb,aps,pre]{revtex4}
\usepackage{graphicx}
\usepackage{dcolumn}
\usepackage{bm}
\begin{document}
\title{Entropy in the natural time-domain\footnote{Published in Physical Review E {\bf70}, 011106 (2004).}}
\author{P. A. Varotsos}
\email{pvaro@otenet.gr}
\affiliation{Solid State Section, Physics Department, University of Athens, Panepistimiopolis, Zografos 157 84, Athens, Greece}
\affiliation{Solid Earth Physics Institute, Physics Department, University of Athens, Panepistimiopolis, Zografos 157 84, Athens, Greece}
\author{N. V. Sarlis}
\affiliation{Solid State Section, Physics Department, University of Athens, Panepistimiopolis, Zografos 157 84, Athens, Greece}
\author{E. S. Skordas}
\affiliation{Solid Earth Physics Institute, Physics Department, University of Athens, Panepistimiopolis, Zografos 157 84, Athens, Greece}
\author{M. S. Lazaridou}
\affiliation{Solid State Section, Physics Department, University of Athens, Panepistimiopolis, Zografos 157 84, Athens, Greece}
\begin{abstract}
 A surrogate data analysis is presented, which is based on the fluctuations of the ``entropy'' $S$
defined in the natural time-domain [Phys. Rev. E {\bf 68}, 031106, 2003]. This entropy is not a static one as, for example, the Shannon entropy. The analysis is applied to three 
types of time-series, i.e., seismic electric signals, ``artificial'' noises and 
electrocardiograms, and ``recognizes'' the non-Markovianity in all these signals. 
Furthermore, it differentiates the electrocardiograms 
of healthy humans from those of the sudden cardiac  death ones. If $\delta S$ and $\delta S_{shuf}$
denote the standard deviation when calculating the entropy by means of a time-window sweeping 
through the original data and the ``shuffled'' (randomized) data, respectively, it seems that the ratio 
$\delta S_{shuf}/\delta S$ plays a key-role. The physical meaning of $\delta S_{shuf}$ is
investigated.
\end{abstract}
\maketitle

\section {introduction}\label{sec1}

In an electric signal consisting of N pulses, the natural time was introduced\cite{var01,var02} 
by ascribing to the $k$-th pulse the value $\chi_{k} = k/N$. The analysis is then made in terms of 
the couple ($\chi_{k}$, $ Q_{k}$) where $ Q_{k}$ stands for the duration of the $k$-th pulse.
The entropy $S$, defined\cite{var01, var03b}  as 
$S \equiv \langle \chi \ln \chi \rangle -\langle \chi \rangle 
\ln \langle \chi \rangle$,
where $\langle \chi \rangle = \sum_{k=1}^N p_k \chi_k$, $p_k$=$Q_{k}/\sum_{n=1}^N Q_n$ and 
$\langle \chi \ln \chi \rangle = \sum_{k=1}^N p_k \chi_k \ln \chi_k$,
was found\cite{var03b} to distinguish Seismic Electric Signals (SES) 
activities from artificial noises (AN), where the latter terminology stands for 
electrical disturbances which are recorded at a measuring site due to 
nearby man-made electric sources. More precisely, SES activities 
and AN have $S$-values smaller and larger than that
 ($S_{u}$) of a ``uniform'' (u)
distribution, respectively (as the latter was defined in 
Refs. \cite{var01,var03,var03b}). Furthermore, ion 
current fluctuations in membrane channels  (ICFMC) have $S$ very close to $S_{u}$ \cite{var03b}. 

The fact that a system contains nonlinear components does not necessarily reflect that a 
specific signal we measure from the system also exhibits nonlinear features. Thus, before 
analyzing this signal by applying nonlinear techniques, we must first clarify if the use of such 
techniques is justified by the data available. The method of surrogate data has been 
extensively used to serve such a purpose (see Ref. \cite{sch00} for a review). Surrogate data refer 
to data that preserve certain linear statistic properties of the experimental data, but 
are random otherwise\cite{cha95,siw01}. These data are prepared by various procedures; 
for example, Siwy et al.\cite{siw01} in order to study the nature of dwell-time series in ICFMC, among other methods, also used 
surrogate data which have been obtained by three  different procedures. 
 The present paper aims, in general, 
at presenting a kind of surrogate data analysis using the entropy fluctuations in the natural 
time-domain (see below) as discriminating statistics. Throughout the paper, the surrogate data 
are obtained by  shuffling the $Q_k$ randomly and hence their 
distribution is conserved. Applying such a procedure, we do the following: consider the
 null hypothesis that the data consist of 
{\em independent} draws from a fixed probability distribution of the dwell times;
 if we find significantly 
different serial correlations in the data and their shuffles, we can reject the hypothesis of 
{\em independence}, see paragraph 3.1 
of Ref. \cite{sch00}. In other words, the tested null hypothesis is that $Q_k$ 
are independent and identically distributed (iid) random variables, i.e., that there are no 
correlations between the lengths of consecutive intervals. If the original (continuous) time series
is Markovian then the null hypothesis for the $Q_k$ should hold, i.e., the $Q_k$ are iid. We
emphasize that the terminology ``Markovian'' throughout this paper always refers to the 
original time series.

Here, as a measure of the natural time entropy fluctuations we consider 
the standard deviation $\delta S$ 
when we calculate the value of $S$ for a number of consecutive pulses 
and study how $S$ varies when 
sweeping this time-window through the whole time-series. 
We use the following three data sets: Two of them  
are those treated in Ref. \cite{var03b}, i.e., 
SES activities and AN. As a third one, we preferred to use, 
instead of ICFMC, the case of electrocardiograms (ECG), for several reasons, 
chief among of which are: (a) They are publicly accessible \cite{GOL00}. 
(b) Instead of the single ICFMC example, a large variety of ECG are available 
(i.e., 105 individuals are employed here, 10 healthy and 95 patients). 
(c) The case of ECG is similar to ICFMC, in the sense that the $S$-value in ECG 
results very close to $S_{u}$ as in ICFMC investigated in \cite{var03b}. 
Note, however, that the intervals
 between heart beats fluctuate widely, e.g., \cite{chia02}.

A general agreement about whether or not normal heart dynamics are chaotic 
or not chaotic is still lacking (e.g., see Ref.\cite{p2} and references therein). 
 The most commonly used non-linear complexity measures are fractal dimensions of
 various kinds (e.g., correlation dimension, Renyi dimensions). Each of them measures 
different aspects of the {\em statistics} on the attractor. On the other hand, 
Liapunov exponents and the Kolmogorov-Sinai entropy (K-S entropy) and entropy rates 
are measures of the {\em dynamics} on an attractor. Except for the K-S entropy and entropy rates,
 the other categories of complexity measures assume a purely 
deterministic system (e.g., see Ref.\cite{p11}). Since a physiological time series may be due to a mixed process, 
stochastic and deterministic, the use of fractal dimensions in physiological time 
series has been occasionally criticized\cite{p11}. On the other hand, entropy is a 
concept equally applicable to deterministic as well as stochastic processes.
This is why we preferred to use the entropy in natural time (more
precisely its fluctuations $\delta S$ ) as  discriminating statistics. 
The following point, 
however, should be stressed. Complexity measures based on {\em static} entropy 
(e.g., Shannon entropy) quantify {\em statistical} order in the time series. 
The underlying key-property of these complexity measures is the probability
 distribution of the (dwell times in the) data analyzed; thus, the result of such computations should 
be independent of permutations performed on the (sequence of the dwell times in the) 
time series as in a surrogate (randomized)
 data set obtained by data shuffling. On the other hand, the entropy in 
natural time (and the relevant measures) considers, from its definition, the {\em sequential} 
order (of beats); in other words, $S$ is a {\em dynamic} entropy, i.e., it captures 
characteristics of the dynamics in a system. Additional comments on the importance 
of the fluctuations of $S$ in ECG will be forwarded in Section \ref{sec5}.

In all examples, we use a sliding window of length three to ten pulses, 
except otherwise stated. Concerning the symbols: We reserve $\delta S$ {\em only} 
for the case when the calculation is made by a {\em single} time-window,
 e.g., 5 pulses. The symbol $\overline{ \delta S}$ denotes 
the average of the $\delta S$-values calculated for a sequence of single windows, 
e.g., 3, 4 and 5 pulses. Finally, $\langle \delta S \rangle$ stands for the  
$\delta S$-values averaged over a group of individuals, e.g., the 10 healthy subjects. 

 The present paper is organized as follows: In section \ref{sec2}, we investigate whether 
a distinction between SES activities and AN can be achieved by the 
$\delta S$-value alone. 
Furthermore, we examine if $\delta S$ can recognize the non-Markovianity in all the signals investigated. 
In section \ref{sec3}, we attempt to shed light on the quantity $\delta S_{shuf}$ 
calculated in a surrogate (randomized) data set obtained by data ``shuffling''. 
We find that $\delta S_{shuf}$ in ECG is a measure of 
$\sigma / \mu$ (where $\mu$ and $\sigma$ stand for the mean value and
 the standard deviation 
of the corresponding intervals, see below). Section \ref{sec4} shows that the $\delta S_{shuf}$-value 
differs from $\delta S$, as expected 
(cf. the entropy $S$  is {\em not} 
static entropy,as mentioned above). 
The prominent role of the ratio $\delta S_{shuf}/\delta S$ in distinguishing ECG of healthy humans 
from those suffered from sudden cardiac death is shown in Section \ref{sec5}.
The conclusions are summarized in Section \ref{sec6}. 
Finally, an Appendix is reserved to derive an exact 
relationship between $\delta S_{shuf}$ and $\sigma / \mu$   when $Q_k$ are iid.

\section{the possibility of employing $\delta S$ to ``recognize'' the non-Markovianity} \label{sec2}
\begin{figure}
\includegraphics{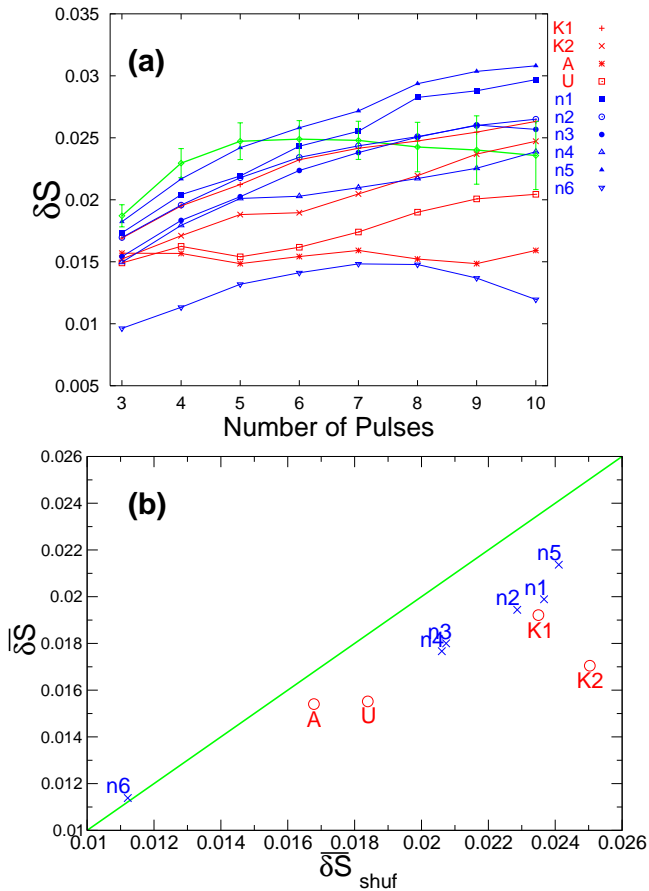}
\caption{\label{fig1} (Color) (a) The $\delta S$-values for 
each SES-activity and artificial 
noise versus the time-window length. The corresponding 
values for a Markovian 
time-series ($10^2$ pulses) are also plotted (green). 
(b) $\overline{ \delta S}$ versus $\overline{ \delta S}_{shuf}$(time-window range 3-5) 
for all the SES 
activities and AN in (a). The straight line corresponds to 
$\overline{ \delta S}_{shuf}$ = $\overline{ \delta S}$.}
\end{figure}

We start by examining whether the $\delta S$-values alone can distinguish 
SES activities from AN as well as ``recognize'' their non-Markovianity. Recall\cite{var02,var03b},
 that SES and AN are time-series of dichotomous nature which are non Markovian. 
In a {\em dichotomous} Markovian time-series, the 
dwell times ($Q_k$) are exponentially distributed; for such a series we plot,  
in Fig. \ref{fig1}(a),
  the $\delta S$-value versus the time-window length. (Since in the 
calculation of $S$  
only ratios of  $Q_k$ are involved the result does not depend on the transition 
rates of the Markovian process.)  
The error shown in this case is on the average 7\%. (The calculation 
was made for a total number of $10^2$ pulses, see below. Note that this error 
decreases upon increasing the number of pulses, i.e., it becomes $\approx$ 2\% for $10^3$ pulses, 
which will be used later). 
In the same figure, we insert the $\delta S$-values calculated for the four SES 
activities (labelled K1, K2, A, U) and the six AN (labelled n1 to n6) depicted in Fig. 1 
of Ref. \cite{var03b}. An inspection of Fig.\ref{fig1}(a) reveals the following conclusions. 
First, {\em no} distinction between SES activities and AN (both of which have estimation 
errors comparable to the aforementioned error of the Markovian) is obvious. 
An inspection of Table I of \cite{EPAPS03},
reveals that the number of pulses in three (out of the four) 
SES activities is around $10^2$, for K2, U 
and A (while for K1, is $\approx 310$) and this is why we calculated here the 
Markovian case for $10^2$ pulses. Second, concerning the possibility of 
``recognizing'' the non-Markovianity (as discussed and shown in 
Refs. \cite{var02,var03,var03b} 
by independent procedures):
 This could be possibly supported,
 {\em only} for the shorter 
time-windows  (i.e., 3, 4 and possibly 5 pulses) for all SES 
activities as well as for most 
AN (i.e., n6, n4, n3, n2, possibly n1, but {\em not} for n5), 
see Fig.\ref{fig1}(a).

\begin{figure}
\includegraphics{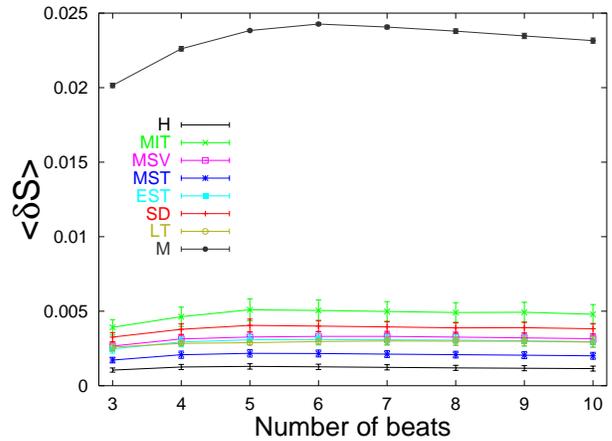}
\caption{\label{fig2}  (Color) The $\langle \delta S \rangle$-values for the QRS-intervals (see the text) of the seven groups 
of humans versus the time-window length. The corresponding values for a 
Markovian time-series ($10^3$ pulses, labelled M) are also plotted. }
\end{figure}

We now investigate if the $\delta S$-values alone can ``recognize'' 
the non-Markovianity in ECG. In a single sinus (normal) cycle of an ECG, 
the turning points are labelled with the letters P, Q, R, S and T. We used here the 
QT database from physiobank \cite{GOL00}( see also \cite{LAG97}), which consists 
of 105 fifteen-minute excerpts of Holter recordings as follows: 10 from MIT-BIH 
Normal Sinus Rhythm Database (i.e., healthy subjects, hereafter labelled H), 15 from MIT-BIH Arrhythmia Database (MIT), 
13 from MIT-BIH Supraventricular Arrhythmia Database (MSV), 6 from MIT-BIH ST 
Change Database (MST), 33 from the European ST-T Database (EST), 4 from MIT-BIH 
Long-Term ECG Database (LT) and 24 from sudden death patients from BIH(SD). 
(cf. BIH denotes the Beth Israel Hospital). In Fig. \ref{fig2}, we plot, 
for the 
QRS-interval time-series, the $\delta S$-value averaged over each of the aforementioned 
seven groups versus the time-window length. Since all time-series 
of these seven groups have 
$\approx 10^3$ intervals, we insert in the same figure the results calculated 
for a Markovian case (cf. with the procedure mentioned in the previous 
paragraph) of comparable length $\approx 10^3$.  
We see that the Markovian case exhibits $\delta S$-values that are roughly one 
order of magnitude larger than those of the seven groups of humans, which clearly 
points to the non-Markovianity of {\em all} the signals in these groups.  We emphasize that the same 
conclusions are drawn if we 
consider, instead of QRS-, the series of QT-intervals, or the  beat-to-beat 
intervals (RR). 
In summary, the $\delta S$-value alone can well recognize
 the non-Markovianity in ECG. 

\begin{figure}[b]
\includegraphics{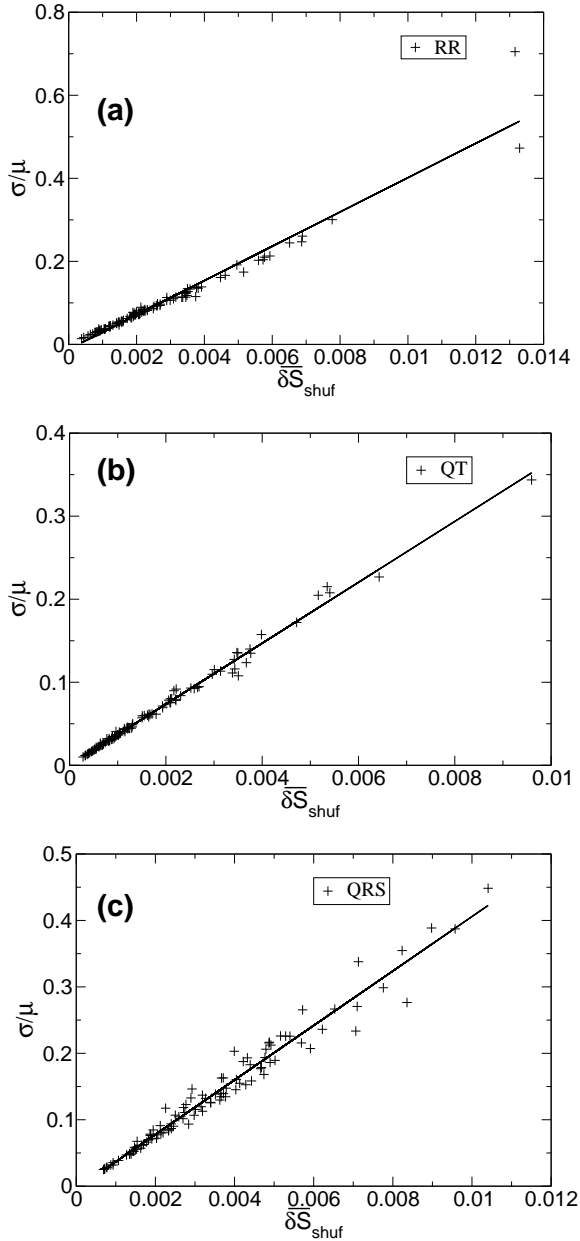}
\caption{\label{fig3}The $\sigma / \mu$-value, for each of the 105 individuals, versus 
the corresponding $\overline{ \delta S}_{shuf}$-value 
for the (a) RR-, (b) QT- and (c) QRS-intervals. The identity of the individual 
associated with each point can be found in Ref.\cite{EPAPS}.
}
\end{figure}

\begin{figure}[b]
\includegraphics{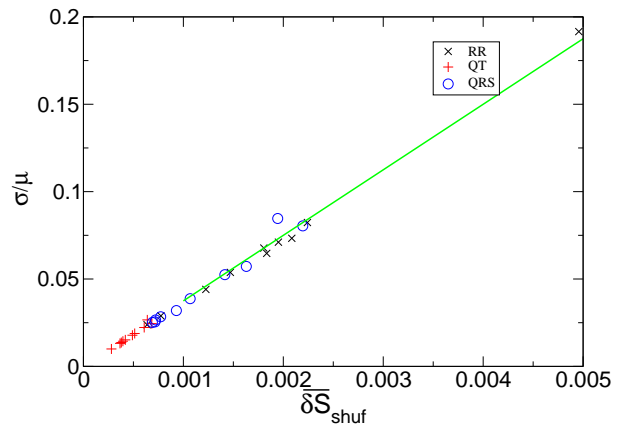}
\caption{\label{fig4} (Color) The $\sigma/\mu$-value for  RR-, QT- and 
QRS-intervals of the ten H 
versus the corresponding 
$\overline{ \delta S}_{shuf}$-value (time-window range 3-10 beats). 
The straight 
line results from a least squares fit of all the thirty points. 
For the identity of the individual associated with each point 
see Ref. \cite{EPAPS}. }
\end{figure}

\begin{figure}[b]
\includegraphics{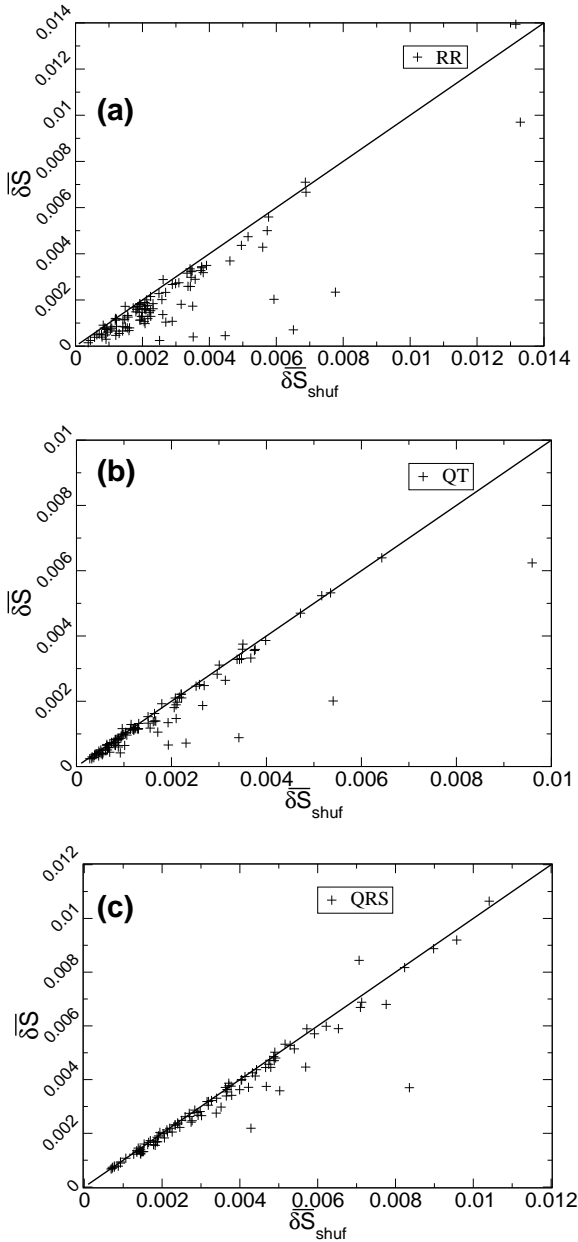}
\caption{\label{fig5}The $\overline{ \delta S}$-value, 
for each of the 105 individuals versus the corresponding $\overline{ \delta S}_{shuf}$-value
for (a) RR-, (b) QT- and (c) QRS-intervals. The straight line, drawn in each case,
corresponds to $\overline{ \delta S}_{shuf}$ = $\overline{ \delta S}$.
For the identity of the individual associated with each point 
see Ref. \cite{EPAPS}.}
\end{figure}

\begin{figure}
\includegraphics{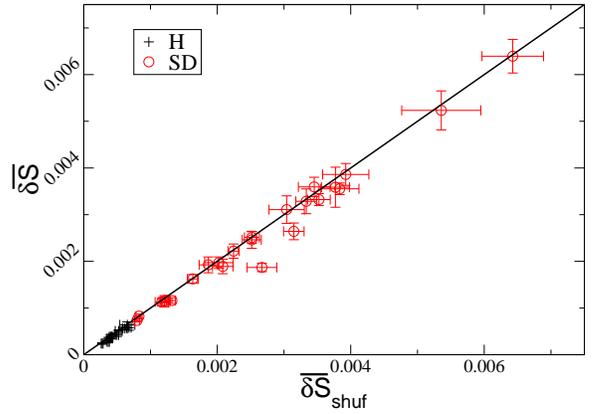}
\caption{\label{fig6} (Color)
The $\overline{ \delta S}$-value,
in each of the 10 H (black) and 24 SD (red), for the QT-intervals versus $\overline{ \delta S}_{shuf}$ 
(time-window range: 3-10 beats). Note that the values of the ordinates are
 appreciably smaller than the $\delta S$-value ($\approx 2\times 10^{-2}$) of the Markovian time-series 
($10^3$ events) depicted in Fig. \ref{fig2}.}
\end{figure}

\section{the physical meaning of $\delta S_{shuf}$} \label{sec3}

In Fig. \ref{fig3}(a) we plot, for each of the 105 individuals, the value 
of $\sigma / \mu$ versus the corresponding value of $\overline{ \delta S}_{shuf}$ 
(time-window range 3-10 beats) for the RR-intervals. 
The same is repeated in Figs. \ref{fig3}(b) and \ref{fig3}(c) for the QT- and QRS-intervals, 
respectively. All these three plots, can be described by linear behavior 
and a least squares fitting to a straight line passing through the origin 
leads to the following slopes: 
38.6 $\pm$ 0.6, 36.8 $\pm$ 0.2 and 40.1 $\pm$ 0.4, for the RR-, QT- and QRS-intervals, 
respectively.
 This points to the conclusion that 
$\delta S_{shuf}$ provides, as intuitively expected, a measure of $\sigma/\mu$. 
(This, however, {\em cannot}  be supported with certainty for 
the SES activities and AN.) 
Note that, although these three slopes are more or less comparable, 
they differ by 
amounts lying outside their standard error. 
Furthermore, it may be worthwhile to 
mention, that if we study {\em altogether} the RR-, QT- and QRS-intervals,
  for 
the 10 healthy humans {\em only} (Fig. \ref{fig4}),
 a good linearity of $\sigma/\mu$ 
versus $\overline{ \delta S}_{shuf}$ results with a slope 37.5 $\pm$ 0.4. 
(cf. if we study each of the three intervals separately, we find slopes that agree 
within the error margins, i.e., 37.5 $\pm$ 0.4, 37.1 $\pm$ 0.7 and 37.8 $\pm$ 0.1 for 
the RR-, QT- and QRS-intervals, respectively). The origin of this {\em common} 
behavior merits further investigation.

One could argue that $Q_k$ may become iid upon their shuffling.
 In the Appendix, we show that,  when 
$Q_k$ are iid, $\delta S$ is  actually  proportional to $\sigma/\mu$; 
the following relationship is obtained:
\begin{eqnarray}
\delta S_{shuf} &=& \frac{\sigma}{\mu} \frac{1}{\sqrt{N-1}}  [  
\sum_{k=1}^N \left(\frac{k}{N}\ln \frac{k}{e N  \overline{\chi}  } \right)^2 \frac{1}{N} \nonumber \\
& -& \left( \sum_{k=1}^N \frac{k}{N^2} \ln \frac{k}{e N \overline{\chi}}   \right)^2 
]
\label{eq1}
\end{eqnarray}
where 
\begin{equation}
\label{eq2}
\overline{\chi}  = \sum_{k=1}^N \frac{k}{N^2}   = \frac {1} {2}+ \frac {1} {2N}
\end{equation}
and $e$ denotes, as usually, the base of the natural logarithms.
The relation (\ref{eq1})  reveals  that $\delta S_{shuf}$ versus $\sigma/\mu$ must be 
a straight line with a slope  ranging from 34.2 to 40.4, for a time-window length 3 to 10. 
This result is comparable with the slopes determined above from the analysis of the ECG data.

\section{on the difference between $\delta S$ and $\delta S_{shuf}$} \label{sec4}

We first comment on the difference between $\delta S$ and $\delta S_{shuf}$ in the SES activities and AN. In Fig. \ref{fig1}(b), the value of 
$\overline{ \delta S}$ versus the corresponding $\overline{ \delta S}_{shuf}$ was plotted, 
for each of the ten signals discussed in Fig. \ref{fig1}(a). The average values 
in Fig. \ref{fig1}(b) have been calculated over the three time-windows 
of 3, 4 and 5 pulses, 
since we mentioned in Section \ref{sec2}  that the ``recognition'' of the non-Markovianity  in all 
SES activities becomes possible  in this time-window range. 
If we disregard n6, 
and despite the errors of around 5\% (for the time-window range 3-5), we may say 
that there is a systematic tendency pointing to a value of 
$\overline{ \delta S}_{shuf}/\overline{ \delta S}$ larger than unity (cf. 
the same 
conclusion is drawn, if we take the averages over the time-window 
range 3-10). 
This is consistent with the non-Markovianity   of all these signals, 
because for a Markovian case we expect 
$\overline{ \delta S}_{shuf}$=$\overline{ \delta S}$. (Since, by definition, 
$\delta S_{shuf}$ corresponds to the entropy fluctuations upon {\em random} mixing 
of $Q_k$, see Section \ref{sec1}, it is naturally expected that in 
a Markovian case the two quantities 
$\delta S$ and $\delta S_{shuf}$ {\em should} coincide). Note that the reverse is {\em not} 
always true (thus the equality $\overline{ \delta S}_{shuf}$ =  $\overline{ \delta S}$ may 
also hold for {\em non}-Markovian time series) as  will be demonstrated  below 
with precise examples. 

We now proceed to compare $\delta S_{shuf}$ with $\delta S$ in ECG. 
Figure \ref{fig5}(a) depicts the $\overline{ \delta S}$-values, calculated for each 
of the 105 individuals, versus the corresponding $\overline{ \delta S}_{shuf}$ for 
the RR-intervals (time-window range 3-10 beats).
The same is repeated in Figs. \ref{fig5}(b) and \ref{fig5}(c) for 
the QT- and QRS-intervals, respectively. In each case, we also plot the straight 
line $\overline{ \delta S}_{shuf}$=$\overline{ \delta S}$ to visualize that the vast 
majority of points fall below this line. The non-equality of $\overline{ \delta S}_{shuf}$ 
and $\overline{ \delta S}$ has been also verified by applying the Wilcoxon paired signed-rank 
test recommended\cite{mot95} to be followed for non-Gaussian paired data. The tested null hypothesis is that the means of $\overline{ \delta S}_{shuf}$
and $\overline{ \delta S}$ are the same and is rejected at a level of significance well below 0.01,
 since the data of Figs. 5(a),(b) and (c) lead to normally distributed
 variables $z=-8.29, -6.81$ and $-6.32$,
 respectively (cf. the corresponding one-tailed asymptotic significance is given by $P(Z<z)$, 
i.e.,  the probability to obtain a normally distributed variable which is smaller than $z$).
 Note that a least squares fit to a straight line 
passing through the origin, results  in the following expressions: 
$\delta S=(0.76\pm 0.03)\delta S_{shuf}$, $\delta S=(0.85\pm 0.02)\delta S_{shuf}$, 
$\delta S=(0.94\pm 0.02)\delta S_{shuf}$ for the Figs. \ref{fig5}(a),\ref{fig5}(b),
\ref{fig5}(c), respectively. The sampling rate $F_s$ in ECG is 250 Hz; thus, if we take 
as an example the RR-intervals,  the experimental error in their allocation is around $1/F_s$. 
The latter reflects in the calculation of $\delta S$ and $\delta S_{shuf}$ errors 
which are drastically smaller than those required to eventually justify a compatibility of 
the expression $\overline{ \delta S} = (0.76\pm 0.03)\overline{ \delta S}_{shuf}$, 
found from Fig.\ref{fig5}(a), with a straight line of slope equal to unity,
 i.e., $\overline{ \delta S} = \overline{ \delta S}_{shuf}$.

The difference between $\delta S$ and 
$\delta S_{shuf}$ in ECG could be understood in the context that the former depends on 
the {\em sequential} order (of beats), as mentioned in Section \ref{sec1}, while the latter does not. Since 
short- (and long-) range correlations is a usual 
feature( see Ref. \cite{PNAS02} and references
 therein)
 in heartbeat dynamics, which are possibly destroyed (or become
weaker) upon randomizing the data, more
``disorder'' is intuitively expected to appear after randomization, thus reflecting
$\delta S_{shuf} > \delta S$. 
Furthermore, note that in {\em all} the three plots of Fig. \ref{fig5} there are some 
drastic deviations from the straight line $\overline{ \delta S} = \overline{ \delta S}_{shuf}$. 
The origin of these deviations is currently investigated in detail.

Finally, we further clarify the aforementioned point
 that the equality $\overline{ \delta S} = \overline{ \delta S}_{shuf}$ does {\em not} 
necessarily reflect Markovianity. In Fig.\ref{fig6}, we plot, for the QT-intervals, 
$\overline{ \delta S}_{shuf}$ versus $\overline{ \delta S}$ (for time-window range 3-10 beats) for SD and H. 
We see that there are several individuals (mainly SD, see also next section) of which 
their points lie practically (i.e., within the error margins) 
on the straight line $\overline{ \delta S} = \overline{ \delta S}_{shuf}$. 
If we plot their $\delta S$- (or $\delta S_{shuf}$-) values versus the time-window 
(in a similar fashion as in Fig. \ref{fig2}), we find that these values are distinctly smaller
 than those of the Markovian case, thus making clear that these
 individuals cannot be characterized as exhibiting Markovian behavior. 
(This non-Markovianity holds for {\em all} H and {\em all} SD.)

\section{the use of $\overline{ \delta S}_{shuf} / \overline{ \delta S} $ to 
distinguish ECG of healthy humans from the sudden cardiac death ones} \label{sec5}

Here we focus only on two groups of ECG, namely H and SD, and examine whether 
they can be distinguished by means of the ratio $\overline{ \delta S}_{shuf} / \overline{ \delta S} $. 
We calculate this ratio, for each 
type of interval, at two ranges: (i) a short (s) range 3-4 beats (cf. consider that the {\em smallest} 
number allowed for the natural time-domain analysis is 3 beats) and (ii) a longer (L) range 50-70 beats. 
For the sake of convenience, we define $\nu \equiv \overline{ \delta S}_{shuf} / \overline{ \delta S} $, 
and hence the following 
ratios are investigated: $\nu_s(\tau)$ and $\nu_L(\tau)$, where $\tau$ denotes the type of 
interval (i.e., $\tau$=RR, QRS or QT) and $s, L$ refer to the range studied (i.e., $s=$ 3-4 
beats and $L$=50-70 beats).

The calculated values for $\nu_s(\tau)$ and $\nu_L(\tau)$ for the three types of intervals are given, for all 
H and SD, in Table \ref{tab1}. The minima $\min_H[\nu_\kappa (\tau)]$ and maxima $\max_H[\nu_\kappa (\tau)]$ (where 
$\kappa$ denotes either the short, $\kappa=s$, or the longer, $\kappa=L$, range) among the healthy subjects are also inserted in 
two separate rows, for each type of interval and each range studied. These minima and maxima 
are labeled $H_{min}$ and $H_{max}$, respectively. The cases of SD which have smaller and larger values 
than $H_{min}$ and $H_{max}$ (reported in each column) are marked with superscripts ``a'' and ``b'' respectively.

A careful inspection of Table \ref{tab1} leads to the following main conclusion: {\em All} SD violate one or more H-limits 
(i.e., they have values that are smaller than $H_{min}$ or larger than  $H_{max}$). We intentionally emphasize that this 
conclusion is also drawn  {\em even} when disregarding the results for the QT-intervals. (Concerning the latter intervals: 
Only 5 SD out of 24 violate the H-limits; however,
 in {\em all} SD, their $\delta S$-values themselves are larger than those in H,
 see also Fig. \ref{fig6}. The usefulness of this difference will be discussed in 
detail elsewhere). In other words, when focusing our investigation solely on the RR-
 and QRS-intervals, {\em all} SD violate one or more of the four H-limits 
related to $\nu_s(RR)$, $\nu_L(RR)$, $\nu_s(QRS)$ and $\nu_L(QRS)$. This is important
 from practical point of view, because the RR- and QRS-intervals can be 
detected more easily (and accurately) than the QT- by means of an automatic
 threshold based detector (e.g., see Ref. \cite{jan97}, which evaluated the results 
of a detector that has been forwarded in Refs. \cite{lag90} and \cite{lag94} to 
determine automatically the waveform limits in Holter ECG).

\squeezetable
\begin{table*} 
\caption{ \label{tab1}The values of the ratios $\overline{\delta
S}_{shuf}/ \overline{{\delta S}}$ in the short ($s$) range 3-4 ($\nu_s$) or in the longer ($L$) range 
50-70 beats ($\nu_L$) in H (sel16265 to sel17453) and SD (sel30 to sel17152) for the RR-, 
QRS- and QT-intervals} 
\begin{ruledtabular}
\begin{tabular}{c ccc ccc}

  individual & 3-4 beats ($\nu_s$) & & & 50-70 beats ($\nu_L$) & 
\\

 \cline{2-4}
 \cline{5-7}
\\
  & \vspace{0.05cm} RR & QRS & QT & RR & QRS & QT \\
 \hline 
 sel16265 & 1.82 & 1.00 & 1.24 & 0.48 & 1.02 & 0.76 \\
 sel16272 & 1.74 & 0.99 & 0.98 & 0.77 & 1.08 & 1.11 \\
 sel16273 & 2.21 & 1.00 & 1.48 & 0.50 & 0.88 & 0.71 \\
 sel16420 & 1.55 & 0.98 & 1.08 & 0.53 & 1.09 & 0.90 \\
 sel16483 & 2.25 & 1.02 & 1.14 & 0.52 & 1.16 & 0.92 \\
 sel16539 & 1.42 & 1.06 & 1.25 & 0.50 & 1.08 & 0.65 \\
 sel16773 & 1.94 & 1.00 & 0.99 & 0.44 & 1.05 & 0.96 \\
 sel16786 & 1.42 & 1.00 & 1.19 & 0.56 & 1.04 & 0.77 \\
 sel16795 & 1.18 & 0.98 & 1.08 & 0.73 & 0.96 & 0.99 \\
 sel17453 & 1.38 & 1.01 & 1.02 & 0.56 & 0.98 & 0.81 \\
\\
 $H_{min}$ & 1.18 & 0.98 & 0.98 & 0.44 & 0.88 & 0.65 \\
 $H_{max}$ & 2.25 & 1.06 & 1.48 & 0.77 & 1.16 & 1.11 \\
\\
sel30 & 1.29 & 1.11\footnotemark[2] & 1.09 & 0.65 & 0.72\footnotemark[1] & 1.09 \\
sel31 & 0.96\footnotemark[1] & 1.08\footnotemark[2] & 1.17 & 1.23\footnotemark[2] & 0.94 & 0.62\footnotemark[1] \\
sel32	& 1.39   & 1.14\footnotemark[2]  &  1.12  & 1.02\footnotemark[2]   &  0.69\footnotemark[1]   & 0.90 \\
sel33	& 1.05\footnotemark[1] & 0.99 & 1.00 & 0.86\footnotemark[2] & 0.82\footnotemark[1] & 0.99 \\
sel34	& 2.11   & 1.29\footnotemark[2]  & 1.11  &  0.42\footnotemark[1] &	 0.78\footnotemark[1]   &  0.67 \\
sel35	& 1.00\footnotemark[1] &  1.00  &  0.96\footnotemark[1]  &  1.01\footnotemark[2]   &   1.05   &  	 1.08 \\
sel36	& 1.02\footnotemark[1] & 1.02  & 1.04  &  0.92\footnotemark[2]   &  1.00   &  0.88 \\
sel37	& 1.07\footnotemark[1] & 1.18\footnotemark[2]  & 1.07  &  0.55   & 0.75\footnotemark[1]  & 0.65 \\
sel38	& 0.99\footnotemark[1] & 1.09\footnotemark[2]  &  1.13  & 1.37\footnotemark[2] & 0.89 & 1.04 \\
sel39	& 0.96\footnotemark[1] & 1.02  & 1.06 &  2.93\footnotemark[2] & 0.92 & 0.90 \\
sel40	& 1.01\footnotemark[1] & 1.00  & 0.93\footnotemark[1] & 0.78\footnotemark[2] & 0.93 & 1.29\footnotemark[2] \\
sel41	& 1.07\footnotemark[1] &  1.04  & 1.02 & 1.07\footnotemark[2] & 0.84\footnotemark[1] & 0.96 \\
sel42	& 1.63 & 1.08\footnotemark[2] & 1.23 & 0.42\footnotemark[1] & 1.06 & 0.67 \\
sel43	& 2.71\footnotemark[2] & 1.11\footnotemark[2] & 1.05 & 0.56 & 0.76\footnotemark[1] & 0.89 \\
sel44	& 0.91\footnotemark[1] & 0.95\footnotemark[1] & 0.88\footnotemark[1] & 2.24\footnotemark[2] & 1.46\footnotemark[2] &  1.32\footnotemark[2] \\
sel45	& 0.98\footnotemark[1] &  1.24\footnotemark[2]  & 1.29 &  0.98\footnotemark[2] &  	 0.86\footnotemark[1] & 0.79 \\
sel46	& 1.03\footnotemark[1] &  1.01 &  1.03 &  1.00\footnotemark[2] & 0.84\footnotemark[1] &  1.01 \\
sel47	& 1.56   & 0.97\footnotemark[1] & 1.03 &  0.45 & 0.97 &  1.01 \\
sel48	& 0.82\footnotemark[1] & 1.18\footnotemark[2] & 1.44 & 1.48\footnotemark[2] & 0.68\footnotemark[1] & 0.73 \\
sel49	& 0.93\footnotemark[1]  & 1.11\footnotemark[2] & 0.96\footnotemark[1] & 1.22\footnotemark[2] & 0.70\footnotemark[1] &  1.14\footnotemark[2] \\
sel50	& 1.05\footnotemark[1] & 0.98 & 0.98 & 0.93\footnotemark[2] & 1.23\footnotemark[2] & 1.50\footnotemark[2] \\
sel51	& 1.25  &  1.01 & 0.97\footnotemark[1] &  1.05\footnotemark[2] & 1.24\footnotemark[2] & 0.91 \\
sel52	& 1.50  &  1.16\footnotemark[2] &  1.22 & 1.00\footnotemark[2] & 0.73\footnotemark[1] &  0.68 \\
sel17152 & 1.64 &   	 1.01  &  1.04  &  0.90\footnotemark[2]  &  1.01 &	 0.97 \\

\end{tabular}
\footnotetext[1]{These values are smaller than the minimum ($H_{min}$) value of $\overline{\delta
S}_{shuf}/ \overline{{\delta S}}$ in H for each range}
\footnotetext[2]{These values are larger than the maximum ($H_{max}$) value of $\overline{\delta
S}_{shuf}/ \overline{{\delta S}}$ in H for each range} 
\end{ruledtabular}
\end{table*}
 
A further inspection of Table \ref{tab1} leads to the following additional comment: When 
investigating the RR-intervals {\em alone} (which can be detected automatically more easily and 
precisely than the other intervals), i.e., studying $\nu_s(RR)$ and  $\nu_L(RR)$, 
the vast majority of SD (22 out of 24 cases) can
 be distinguished from H (only two SD, i.e., 
sel30 and sel47, obey the corresponding H-limits).
 Specifically, concerning $\nu_s(RR)$, 15 SD have values 
smaller than $H_{min}=1.18$, while only one SD (i.e., sel43) has a value exceeding
 $H_{max}=2.25$; as for $\nu_L(RR)$, 18 SD exceed $H_{max}=0.77$,
 while only 2 SD (i.e., sel34 and sel42) have values smaller than $H_{min}=0.44$.

In what remains, we  proceed to a tentative physical interpretation of the above results, the main 
feature of which focuses on the fact that most SD simultaneously
 have $\nu_s(RR)$-values smaller than $H_{min}(=1.18)$ and $\nu_L(RR)$-values 
exceeding $H_{max}(=0.77)$. The RR time-series 
of healthy subjects are characterized by high 
complexity (e.g., \cite{iva01,PNAS02}); this, if we 
recall that in a Markovian series we intuitively expect $\delta S_{shuf} / \delta S=1$ 
(and hence $\nu_s=1$ and $\nu_L=1$), is compatible with the fact that in 
{\em all} H {\em both} $\nu_s(RR)$ and $\nu_L(RR)$ distinctly differ from unity (see Table \ref{tab1}). 
We now turn to SD by considering that for individuals at 
high risk of sudden death the fractal physiological organization (long range correlations) 
breaks down and this is often accompanied by emergence of {\em uncorrelated randomness} (see 
\cite{PNAS02} and references therein). It is therefore naturally expected that in SD the values of 
$\nu_s(RR)$ and $\nu_L(RR)$ become closer to the Markovian value (i.e., unity) compared to H; thus, in SD, $\nu_s(RR)$ naturally becomes smaller than the value 1.18 (the corresponding $H_{min}$-limit) and $\nu_L(RR)$ larger than 0.77 (the corresponding $H_{max}$-limit).

We now focus on the following important property of H: although {\em both} 
$\nu_s(RR)$ and $\nu_L(RR)$ differ from unity, as mentioned, they systematically behave 
{\em differently}, i.e., $\nu_s(RR) > 1$ while $\nu_L(RR) < 1$. The exact origin of the 
latter difference has not yet been identified with certainty, but the following comments 
might be relevant: First, in the frame of the frequency-domain characteristics of heart 
rate variability (e.g., \cite{sha03}), we may state that $\nu_s(RR)$ and $\nu_L(RR)$ are 
associated with the high-frequency (HF, 0.15-0.4 Hz) and low-frequency (LF, 0.015-0.15 Hz) 
range in the RR tachogram (``instantaneous'' heart rate,1/RR). An important difference on the 
effect of the sympathetic and parasympathetic modulation of the RR-intervals has been 
noticed (e.g., see \cite{sha03} and references therein): Sympathetic tone is believed to 
influence the LF component whereas {\em both} sympathetic and parasympathetic activity 
have an effect on the HF component (recall that our results show $\nu_s(RR) > \nu_L(RR)$). 
Second, at short time scales (high frequencies), it has been suggested \cite{pen95} that 
we have relatively {\em smooth} heartbeat oscillations associated with respiration 
(e.g., 15 breaths per minute corresponds to a 4 sec oscillation with a peak in the power 
spectrum at 0.25 Hz, see \cite{sha03}); this is lost upon randomizing the consecutive 
intervals $Q_k$, thus probably 
leading to (larger variations -compared to the original experimental data- between the durations 
of consecutive intervals and hence to) $\delta S_{shuf}$-values larger than $\delta S$, i.e., 
a $\nu_s(RR)$-value larger than unity (cf. an extension of the current analysis to a surrogate sequence for a simultaneous 
recording of the breath rate and the instantaneous heart rate, upon considering the points  
discussed in paragraph 4.6 of Ref. \cite{sch00}, could greatly 
contribute towards clarifying the validity of such an explanation). Such an argument, if true, 
cannot be applied, of course, in the longer range 50-70 beats and hence explain why the opposite 
behavior, i.e., $\delta S_{shuf}<\delta S$, then holds. The latter finding must be inherently 
connected to the nature itself of the long range correlations. The existence of the latter 
is pointed out from the fact that (in this range also) the RR-intervals result in $\delta S$-values 
($\sim10^{-3}$) which significantly differ from the Markovian $\delta S$-value ($\sim 10^{-2}$) (cf. 
the existence 
of the long range correlations in the heart rate variability has been independently established 
by several applications of the detrended fluctuation analysis, e.g., see \cite{pen95}, \cite{PNAS02} and references therein).

A simplified interpretation of the results of Fig.\ref{fig6}, and in particular the reason 
why for the QT intervals the quantity $\delta S$ is larger for the SD than for the H,
 could be attempted if we consider that: 
(i) $S$ could be thought as a measure of the ``disorder'' (in the consecutive intervals) (ii) the essence 
of the natural time-domain analysis is built on the variation of the durations of consecutive pulses, 
and (iii) it has been clinically observed (e.g., see Ref.\cite{p16}) that the QT interval (which corresponds to 
the time in which the heart in each beat ``recovers'' -electrically speaking- from the previous 
excitation) exhibits frequent prolonged values before cardiac death. 
Thus,  when a time-window is sliding on an H-ECG, it is intuitively 
expected to find, more or less, the same $S$-values (when sweeping through 
various parts of the ECG) and hence a small $\delta S$-value is envisaged. By the same token, 
in an SD-ECG, we expect that, in view of the short-long-short sequences
of the QT-intervals, the corresponding
 $S$-values will be much different (compared to H), thus leading to a larger $\delta S$-value
 (cf. in the 
same frame we may also understand why the $\sigma / \mu$ values -and hence $\delta S_{shuf}$, see 
Eq. (\ref{eq1})- are larger in SD than those in H, as shown in Fig.\ref{fig6}). 
The distinction between SD and H could be also
 understood in the context of dynamic phase-transitions (critical phenomena), as follows:
 In SD, since the dynamic phase transition (cardiac arrest) is approached, the {\em fluctuations}
 of $S$ are expected to become larger, thus reflecting larger $\delta S$; 
such intense fluctuations are not expected, of course, for H.

\section{conclusions} \label{sec6}
The main point emerged in the surrogate data analysis presented in this paper, is the key-role of the 
quantity  $\delta S_{shuf} / \delta S$. This ratio:

1. reveals the non-Markovianity  in all three types of signals analyzed here, i.e., SES activities, 
AN and ECG. In a Markovian case  we have $\delta S_{shuf} = \delta S$, but the reverse is 
not always valid; it may happen that $\delta S_{shuf} / \delta S = 1$, although $\delta S$- 
(and $\delta S_{shuf}$-) value drastically differ from that of the Markovian (this is the case of ECG).

2. differentiates the ECG of healthy humans (H) from those suffered from sudden cardiac death (SD).
More precisely, in SD, the $\delta S_{shuf}/\delta S$ values of the RR (i.e. beat to beat) intervals 
become closer to the Markovian value (i.e., unity) compared to those in H. Furthermore, 
in SD, {\em both} $\delta S$ -and $\delta S_{shuf}$- values of the QT interval (   corresponding to the
time in which the heart ``recovers'' from its previous excitation) are larger than those in H.  

As for the physical meaning of  $\delta S_{shuf}$ in ECG, it was shown to be a measure of $\sigma / \mu$.

\appendix*
\section{Interrelation between $\delta S_{shuf}$ and $\sigma /\mu$ in the case of iid}

 If we consider a time-series $Q_k$, where $Q_k \geq 0, k=1, 2,\ldots N$, we obtain
the quantities $p_k=Q_k/\sum_{l=1}^N Q_l$, which satisfy the necessary conditions\cite{abr70}:
$p_k\geq 0, \sum_{k=1}^N p_k=1$ to be considered as point probabilities. We then define\cite{var01,var02,var03b} the moments of the natural time $\chi_k=k/N$ as $\langle \chi^q \rangle= \sum_{k=1}^N (k/N)^q p_k$ and
the entropy  $S \equiv \langle \chi \ln \chi \rangle -\langle \chi \rangle 
\ln \langle \chi \rangle$, where $\langle \chi \ln \chi \rangle=\sum_{k=1}^N  (k/N) \ln (k/N) p_k$.
This Appendix is solely focused on a uniform distribution in the natural time-domain.

We now consider the case when $Q_k$ are independent and identically distributed (iid) positive random variables. It then follows that the expectation value $\text{E}(p_k)=\text{E} [Q_k/\sum_{l=1}^N Q_l]$ of $p_k$  equals $1/N$:
\begin{equation}
\text{E}(p_k)=\frac{1}{N}.
\label{a1}
\end{equation}
Equation (\ref{a1}) results from the fact that, since $Q_k$ are iid, we have:  
$\text{E}[ \sum_{k=1}^N Q_k/\sum_{l=1}^N Q_l]=1=N \text{E}(p_k)$. 
For the purpose
of our calculations the relation between the variance of $p_k$, 
$\text{Var}(p_k)=\text{E} [ (p_k-1/N)^2]$, and the covariance of $p_k$ and $p_l$,
$\text{Cov} (p_k,p_l)= \text{E}[(p_k-1/N)(p_l-1/N)]$,  is of 
central importance.  Using the constraint 
$\sum_{k=1}^N p_k=1$,  leading to $p_k-1/N=\sum_{l\neq k}(1/N-p_l)$, and the 
fact that $Q_k$ are iid, we obtain 
$\text{E} \left [ (p_k-1/N)^2\right ]=\text{E}\left [(p_k-1/N)\sum_{l\neq k}(1/N-p_l)\right ]=-(N-1)
\text{E}\left[ (p_k-1/N)(p_l-1/N) \right]$. Thus, we get 
\begin{equation}
\text{Cov} (p_k,p_l)=-\frac{ \text{Var}(p_k)}{N-1}.
\label{a2}
\end{equation}
The $N$-dependence of $\text{Var} (p_k)$ is obtained from  
\begin{equation}
\text{Var} (p_k) = \frac{1}{N^2} \text{E} \left[ \left( \frac{N Q_k}{\sum_{l=1}^NQ_l}-1 \right)^2 \right],
\label{a3a}
\end{equation}
where the quantity  $\text{E} [ ( N Q_k/\sum_{l=1}^NQ_l-1)^2 ]$ is {\em asymptotically} $N$-independent. 
The latter arises as follows: If $\text{E} (Q_k)=\mu$ 
and $\text{Var} (Q_k)=\sigma^2 (< \infty)$,  as a result of the central limit 
theorem\cite{fel71}, we have $\text{E}(\sum_{k=1}^NQ_k/N)=\mu$ and
$\text{Var}(\sum_{k=1}^NQ_k/N)=\sigma^2/N$. The latter two equations, for large enough $N$ imply that 
$\text{E} [ ( N Q_k/\sum_{l=1}^NQ_l-1)^2 ]\approx \text{E} [ ( Q_k/\mu-1)^2 ]=\sigma^2/\mu^2$.
Thus, Eq.(\ref{a3a}) becomes
\begin{equation}
\text{Var} (p_k) = \frac{\sigma^2}{N^2\mu^2}. 
\label{a3}
\end{equation} 

We now turn to the statistical properties of $\langle \chi^q \rangle$. Using Eq.(\ref{a1}),
we have
\begin{equation}
\text{E} [ \langle \chi^q \rangle ]=\sum_{k=1}^N \left(\frac{k}{N}\right)^q \frac{1}{N}.
\label{a4}
\end{equation}
which,  since\cite{gra80} $\sum_{k=1}^N k^q=N^{q+1}/(q+1)+N^q/2+o(N^q)$, reveals that
$\text{E} [ \langle \chi^q \rangle ]$  is again asymptotically $N$-independent because it 
approaches the value $1/(q+1)$ with a ``small'' $1/(2N)$ correction. 
The variance 
$\text{Var}[ \langle \chi^q \rangle ][=(\delta \langle \chi^q \rangle)^2]$
\begin{equation}
\text{Var}[ \langle \chi^q \rangle ]=\text{E} \left \{  \left[ \sum_{k=1}^N \left(\frac{k}{N}\right)^q \left( p_k-\frac{1}{N}\right ) \right]^2 \right \},
\label{eqvar}
\end{equation} 
after expanding the square and using Eqs.(\ref{a2}) and (\ref{a3}), becomes:
\begin{eqnarray}
\text{Var}[ &\langle \chi^q \rangle & ]= 
\sum_{k=1}^N\left( \frac{k}{N} \right)^{2q}\frac{\sigma^2}{N^2\mu^2} \nonumber \\
&-& \frac{\sigma^2}{(N-1)N^2\mu^2} \sum_{k=1}^N\left( \frac{k}{N} \right)^{q} \sum_{l=1,l\neq k}^N \left( \frac{l}{N} \right)^{q}.
\label{a5} 
\end{eqnarray}
which, using Eq.(\ref{a4}), finally leads to:
\begin{equation}
\text{Var}[ \langle \chi^q \rangle ]=\frac{\sigma^2}{(N-1)\mu^2} 
\left\{ \text{E} [ \langle \chi^{2q} \rangle ]- \text{E}^2 [ \langle \chi^{q} \rangle ]
\right\}.
\label{fd}
\end{equation}

\begin{figure}[b]
\includegraphics{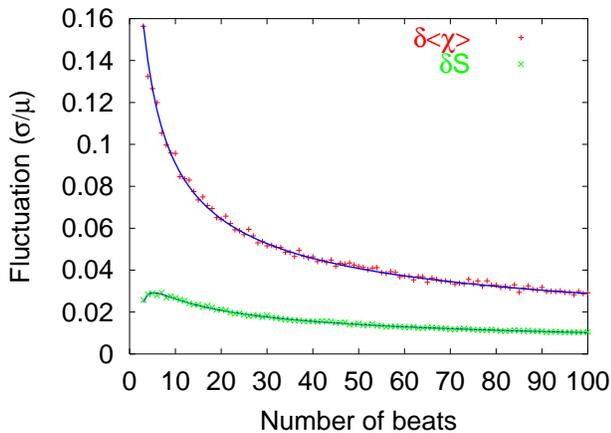}
\caption{\label{FigX} (Color) Comparison of the theoretical estimations (solid lines) of
 $\delta \langle \chi \rangle$ and $\delta S$ resulting from Eqs.(\ref{fd}) and (\ref{ds}), respectively, with the 
values obtained (plus and cross, respectively) using a Gaussian sample having values of $\mu$, $\sigma$ 
and size ($\approx 1000$) similar to those in ECG. Here, as well as throughout 
the paper, the estimator $(\delta X )^2=\sum (X-\overline{X})^2/N$ was used for the calculation 
of the sample variance
of the synthetic data, and thus
$N-1$ was replaced by $N$ in Eqs.(\ref{fd}) and (\ref{ds}).}
\end{figure}

The proof of Eq.(\ref{fd}) can be generalized for all linear functionals of $p_k$ of 
the form $\langle f(\chi) \rangle=\sum_{k=1}^N f(k/N) p_k$ and yields:
\begin{equation}
\text{Var}[ \langle f(\chi) \rangle ]= \frac{\sigma^2 \left\{ \text{E} [ \langle f^2(\chi) \rangle ]- \text{E}^2 [ \langle f(\chi ) \rangle ]\right\}}{(N-1)\mu^2}.
\label{eq8a}
\end{equation}
In Fig.(\ref{FigX}), we compare the theoretical result of Eq.(\ref{fd}) 
with synthetic (Gaussian) data which have values of  $\mu$, $\sigma$ 
and size ($\approx 1000$) similar to those in ECG. Note that when one uses 
the estimator $(\delta X )^2=\sum (X-\overline{X})^2/N$, instead of the unbiased 
estimator  $(\delta X )^2=\sum (X-\overline{X})^2/(N-1)$, in order 
to find the sample variance, $N$ should replace
$N-1$ in Eq.(\ref{fd}).
 
We now proceed to the statistical properties of the entropy\cite{var01,var03b} $S = \langle \chi \ln \chi \rangle -\langle \chi \rangle \ln \langle \chi \rangle$. The expectation value 
\begin{equation}
\text{E}(S)=\text{E} \left[ \sum_{k=1}^N \frac{k}{N} \ln \left( \frac{k}{N} \right) p_k -
\sum_{k=1}^N \frac{k}{N} p_k \ln \left( \sum_{l=1}^N \frac{l}{N} p_l \right) \right]
\end{equation}
can be evaluated as follows: we add and subtract the term $\sum_{k=1}^N \frac{k}{N} p_k \ln \left[ \sum_{l=1}^N \left( \frac{l}{N} \right)\frac{1}{N} \right]$, and then expand 
the resulting term $\ln [1+ \sum_{l=1}^N \frac{l}{N} (p_l-\frac{1}{N})/\sum_{l=1}^N \frac{l}{N^2}]$ to first order in $(p_l-\frac{1}{N})$; finally, using Eq.(\ref{fd}), we obtain
\begin{eqnarray}
\text{E}(S)&=&\sum_{k=1}^N \frac{k}{N^2} \ln \left( \frac{k}{N} \right) -\sum_{k=1}^N \frac{k}{N^2} \ln \left( \sum_{l=1}^N \frac{l}{N^2}  \right) \nonumber \\
&-& \frac{ \sigma^2 \left[ \sum_{k=1}^N k^2/N^3-(\sum_{k=1}^N k/N^2)^2\right]}{(N-1)\mu^2 \sum_{l=1}^N l/N^2}.
\label{as}
\end{eqnarray}
This equation reveals that $\text{E}(S)$  depends slightly on $\sigma/\mu$;
upon increasing $N$ the last term
of Eq.(\ref{as}) decays  
as $1/N$ (cf. the  sums in the numerator and the denominator 
are of the form $E [ \langle \chi^{q} \rangle ]$, for $q$=1 and 2, and 
asymptotically lead to a constant $1/(q+1)$, see the relevant discussion after Eq.(\ref{a4})).

To simplify the calculation of the variance of the entropy $\text{Var}(S)$, 
we  define the two linear functionals
\begin{equation}
m[x_k]=\sum_{k=1}^N \frac{k}{N} x_k,
\label{a11}
\end{equation} 
\begin{equation}
L[x_k,\xi]=\sum_{k=1}^N \frac{k}{N} \ln \left( \frac{k}{\xi N} \right) x_k,
\label{a12}
\end{equation}
and the constant time-series $\mathbb{K}=\{ x_k \} : x_k=1/N,\, k=1, 2, \ldots N $. Note that for 
both functionals $m[x_k]$ and $L[x_k,\xi]$, in view of their linearity, we have
\begin{equation}
\text{E}\left \{m[p_k-\frac{1}{N}]\right \}=\text{E}\left \{L[p_k-\frac{1}{N},\xi]\right \}=0. 
\label{eqa}
\end{equation}
Using Eqs.(\ref{a11}) and (\ref{a12}) the entropy can be written, in compact form, as follows:
\begin{equation}
S=L\left[ p_k,m [p_k] \right],
\label{s0}
\end{equation}
and its expectation value is written as
\begin{equation}
\text{E} (S)=L[\mathbb{K},1]-m[\mathbb{K}] \ln m[\mathbb{K}]-\frac{\sigma^2
 \kappa_{1,u}}{(N-1)\mu^2 m[\mathbb{K}]},
\end{equation} 
where $\kappa_{1,u}=\text{E} [ \langle \chi^{2} \rangle ]- \text{E}^2 [ \langle \chi \rangle ]$.

The variance of the entropy $\text{Var}(S)=(\delta S)^2$
 can then be found by adding 
and subtracting the terms 
$m[p_k] \ln m[\mathbb{K}]$ and $m[p_k-1/N]$
and using the expansion 
$m[p_k]\ln \frac{m[p_k]}{m[\mathbb{K}]}=m[p_k]m[p_k-1/N]/m[\mathbb{K}]$;
this gives: 
\begin{widetext}
\begin{eqnarray}
\text{Var}(S)&=&\text{E} \left\{ \left( L[p_k,1]-m[p_k]\ln m[p_k] -L[\mathbb{K},1]
+m[\mathbb{K}]\ln m[\mathbb{K}] +\frac{\sigma^2 \kappa_{1,u}}{(N-1)\mu^2 m[\mathbb{K}]} \right)^2 \right\}, 
\nonumber \\
&=&\text{E} \left\{ \left( L[p_k-\frac{1}{N},1]-m[p_k]\ln \frac{m[p_k]}{m[\mathbb{K}]}
+m[\frac{1}{N}-p_k]\ln m[\mathbb{K}] +\frac{\sigma^2 \kappa_{1,u}}{(N-1)\mu^2 m[\mathbb{K}]} \right)^2 \right\},  \nonumber \\
&=&\text{E} \left\{ \left( L \left [p_k-\frac{1}{N},m[\mathbb{K}] \right]-\frac{m[p_k]m[p_k-\frac{1}{N}]}{m[\mathbb{K}]} 
+\frac{\sigma^2 \kappa_{1,u}}{(N-1)\mu^2 m[\mathbb{K}]} \right)^2 \right\},  \nonumber \\
&=&\text{E} \left\{ \left( L \left[ p_k-\frac{1}{N},m[\mathbb{K}] \right] -m[p_k-\frac{1}{N}]-\frac{m^2[p_k-\frac{1}{N}]}{m[\mathbb{K}]} 
+\frac{\sigma^2 \kappa_{1,u}}{(N-1)\mu^2 m[\mathbb{K}]} \right)^2 \right\},   \nonumber \\
&=&\text{E} \left\{ \left( L \left[ p_k-\frac{1}{N},m[\mathbb{K}] e \right] -\frac{m^2[p_k-\frac{1}{N}]}{m[\mathbb{K}]} 
+\frac{\sigma^2 \kappa_{1,u}}{(N-1)\mu^2 m[\mathbb{K}]} \right)^2 \right\}. 
\label{a13}
\end{eqnarray}
Expanding the square in Eq.(\ref{a13}), and using Eq.(\ref{eqa}), we find  
\begin{eqnarray}
\text{Var}(S)= \text{E} \Biggl( L^2 \left [p_k-\frac{1}{N},m[\mathbb{K}] e \right ]
&+& 2L \left [p_k-\frac{1}{N},m[\mathbb{K}] e \right ] 
\frac{m^2[p_k-\frac{1}{N}]}{m[\mathbb{K}]} \nonumber \\
&+&\frac{m^4[p_k-\frac{1}{N}]}{m^2[\mathbb{K}]}
- \frac{\sigma^4 \kappa^2_{1,u}}{(N-1)^2\mu^4 m^2[\mathbb{K}]}
\Biggr).\label{a14} 
\end{eqnarray}
{\em If} we assume that the distribution of $Q_k$ is {\em skewnessless}, i.e. $\text{E} [(Q_k-\mu)^3]=0$, 
the expectation value of the second term in Eq.(\ref{a14}) vanishes, 
whereas the third and the fourth terms are of order $1/N^2$ and hence negligible with 
respect to the first term. Thus,
\begin{equation}
\text{Var}(S)=\text{E} \left(  L^2 \left[  p_k-\frac{1}{N}, m[\mathbb{K}] e \right] \right),
\label{eqx}
\end{equation}
which can be explicitly  written as follows 
\begin{equation} 
\text{Var}(S)=
\text{E} \left\{ \left[ \sum_{k=1}^N \frac{k}{N} \ln \left( \frac{k}{m[\mathbb{K}] N e} \right) 
 \left( p_k-\frac{1}{N} \right) 
\right]^2 \right\}.
\label{afa}
\end{equation}
The right side of Eq.(\ref{afa}) becomes similar to Eq.(\ref{eqvar}), if we replace
$\chi^q$ by  $\chi \ln (\chi/m[\mathbb{K}]e)$; thus 
after expanding the square and using Eqs.(\ref{a2}) and (\ref{a3}), we finally obtain
\begin{equation}
\text{Var}(S)=\frac{\sigma^2}{(N-1)\mu^2} \left[ 
 \sum_{k=1}^N  \left( \frac{k}{N} \ln  \frac{N k}{e\sum_{k=1}^N k}  \right)^2
\frac{1}{N}-
\left( \sum_{k=1}^N \frac{k}{N^2} \ln \frac{N k}{e\sum_{k=1}^N k}  \right)^2 
\right] .
\label{ds}
\end{equation}
\end{widetext}
A comparison of Eqs.(\ref{eqx}) and (\ref{s0}) reveals  the following: in order to find the entropy fluctuation $\delta S$,
one simply has to replace in Eq. (\ref{s0}) $m[p_k]$ with $m[\mathbb{K}]e$ and then
directly take its variance according to Eq.(\ref{eq8a}). Equation (\ref{ds}) 
is just Eq.(\ref{eq1}) of the main text.

\bibliographystyle{apsrev}

\begin{thebibliography}{26}
\expandafter\ifx\csname natexlab\endcsname\relax\def\natexlab#1{#1}\fi
\expandafter\ifx\csname bibnamefont\endcsname\relax
  \def\bibnamefont#1{#1}\fi
\expandafter\ifx\csname bibfnamefont\endcsname\relax
  \def\bibfnamefont#1{#1}\fi
\expandafter\ifx\csname citenamefont\endcsname\relax
  \def\citenamefont#1{#1}\fi
\expandafter\ifx\csname url\endcsname\relax
  \def\url#1{\texttt{#1}}\fi
\expandafter\ifx\csname urlprefix\endcsname\relax\def\urlprefix{URL }\fi
\providecommand{\bibinfo}[2]{#2}
\providecommand{\eprint}[2][]{\url{#2}}

\bibitem[{\citenamefont{Varotsos et~al.}(2001)\citenamefont{Varotsos, Sarlis,
  and Skordas}}]{var01}
\bibinfo{author}{\bibfnamefont{P.~A.} \bibnamefont{Varotsos}},
  \bibinfo{author}{\bibfnamefont{N.~V.} \bibnamefont{Sarlis}},
  \bibnamefont{and} \bibinfo{author}{\bibfnamefont{E.~S.}
  \bibnamefont{Skordas}}, \bibinfo{journal}{Practica of Athens Academy}
  \textbf{\bibinfo{volume}{76}}, \bibinfo{pages}{294} (\bibinfo{year}{2001}).

\bibitem[{\citenamefont{Varotsos et~al.}(2002)\citenamefont{Varotsos, Sarlis,
  and Skordas}}]{var02}
\bibinfo{author}{\bibfnamefont{P.~A.} \bibnamefont{Varotsos}},
  \bibinfo{author}{\bibfnamefont{N.~V.} \bibnamefont{Sarlis}},
  \bibnamefont{and} \bibinfo{author}{\bibfnamefont{E.~S.}
  \bibnamefont{Skordas}}, \bibinfo{journal}{Phys. Rev. E}
  \textbf{\bibinfo{volume}{66}}, \bibinfo{pages}{011902}
  (\bibinfo{year}{2002}).

\bibitem[{\citenamefont{Varotsos
  et~al.}(2003{\natexlab{a}})\citenamefont{Varotsos, Sarlis, and
  Skordas}}]{var03b}
\bibinfo{author}{\bibfnamefont{P.~A.} \bibnamefont{Varotsos}},
  \bibinfo{author}{\bibfnamefont{N.~V.} \bibnamefont{Sarlis}},
  \bibnamefont{and} \bibinfo{author}{\bibfnamefont{E.~S.}
  \bibnamefont{Skordas}}, \bibinfo{journal}{Phys. Rev. E}
  \textbf{\bibinfo{volume}{68}}, \bibinfo{pages}{031106}
  (\bibinfo{year}{2003}{\natexlab{a}}).

\bibitem[{\citenamefont{Varotsos
  et~al.}(2003{\natexlab{b}})\citenamefont{Varotsos, Sarlis, and
  Skordas}}]{var03}
\bibinfo{author}{\bibfnamefont{P.~A.} \bibnamefont{Varotsos}},
  \bibinfo{author}{\bibfnamefont{N.~V.} \bibnamefont{Sarlis}},
  \bibnamefont{and} \bibinfo{author}{\bibfnamefont{E.~S.}
  \bibnamefont{Skordas}}, \bibinfo{journal}{Phys. Rev. E}
  \textbf{\bibinfo{volume}{67}}, \bibinfo{pages}{021109}
  (\bibinfo{year}{2003}{\natexlab{b}}).

\bibitem[{\citenamefont{Schreiber and Schmitz}(2000)}]{sch00}
\bibinfo{author}{\bibfnamefont{T.}~\bibnamefont{Schreiber}} \bibnamefont{and}
  \bibinfo{author}{\bibfnamefont{A.}~\bibnamefont{Schmitz}},
  \bibinfo{journal}{Physica D} \textbf{\bibinfo{volume}{142}},
  \bibinfo{pages}{346} (\bibinfo{year}{2000}).

\bibitem[{\citenamefont{Chang et~al.}(1995)\citenamefont{Chang, Sauer, and
  Schiff}}]{cha95}
\bibinfo{author}{\bibfnamefont{T.}~\bibnamefont{Chang}},
  \bibinfo{author}{\bibfnamefont{T.}~\bibnamefont{Sauer}}, \bibnamefont{and}
  \bibinfo{author}{\bibfnamefont{S.~J.} \bibnamefont{Schiff}},
  \bibinfo{journal}{Chaos} \textbf{\bibinfo{volume}{5}}, \bibinfo{pages}{376}
  (\bibinfo{year}{1995}).

\bibitem[{\citenamefont{Siwy et~al.}(2001)\citenamefont{Siwy, Mercik, Weron,
  and Ausloos}}]{siw01}
\bibinfo{author}{\bibfnamefont{Z.}~\bibnamefont{Siwy}},
  \bibinfo{author}{\bibfnamefont{S.}~\bibnamefont{Mercik}},
  \bibinfo{author}{\bibfnamefont{K.}~\bibnamefont{Weron}}, \bibnamefont{and}
  \bibinfo{author}{\bibfnamefont{M.}~\bibnamefont{Ausloos}},
  \bibinfo{journal}{Physica A} \textbf{\bibinfo{volume}{297}},
  \bibinfo{pages}{79} (\bibinfo{year}{2001}).

\bibitem[{\citenamefont{Goldberger et~al.}(2000)\citenamefont{Goldberger,
  Amaral, Glass, Hausdorff, Ivanov, Mark, Mictus, Moody, Peng, and
  Stanley}}]{GOL00}
\bibinfo{author}{\bibfnamefont{A.~L.} \bibnamefont{Goldberger}},
  \bibinfo{author}{\bibfnamefont{L.~A.~N.} \bibnamefont{Amaral}},
  \bibinfo{author}{\bibfnamefont{L.}~\bibnamefont{Glass}},
  \bibinfo{author}{\bibfnamefont{J.~M.} \bibnamefont{Hausdorff}},
  \bibinfo{author}{\bibfnamefont{P.~C.} \bibnamefont{Ivanov}},
  \bibinfo{author}{\bibfnamefont{R.~G.} \bibnamefont{Mark}},
  \bibinfo{author}{\bibfnamefont{J.~E.} \bibnamefont{Mictus}},
  \bibinfo{author}{\bibfnamefont{G.~B.} \bibnamefont{Moody}},
  \bibinfo{author}{\bibfnamefont{C.-K.} \bibnamefont{Peng}}, \bibnamefont{and}
  \bibinfo{author}{\bibfnamefont{H.~E.} \bibnamefont{Stanley}},
  \bibinfo{journal}{Circulation} \textbf{\bibinfo{volume}{101}},
  \bibinfo{pages}{e215 (see also \url{\http://www.physionet.org})}
  (\bibinfo{year}{2000}).

\bibitem[{\citenamefont{Chialvo}(2002)}]{chia02}
\bibinfo{author}{\bibfnamefont{D.~R.} \bibnamefont{Chialvo}},
  \bibinfo{journal}{Nature} \textbf{\bibinfo{volume}{419}},
  \bibinfo{pages}{263} (\bibinfo{year}{2002}).

\bibitem[{\citenamefont{Glass}(2001)}]{p2}
\bibinfo{author}{\bibfnamefont{L.}~\bibnamefont{Glass}},
  \bibinfo{journal}{Nature} \textbf{\bibinfo{volume}{410}},
  \bibinfo{pages}{277} (\bibinfo{year}{2001}).

\bibitem[{\citenamefont{Zebrowski et~al.}(2000)\citenamefont{Zebrowski,
  Poplawska, Baranowski, and Buchner}}]{p11}
\bibinfo{author}{\bibfnamefont{J.~J.} \bibnamefont{Zebrowski}},
  \bibinfo{author}{\bibfnamefont{W.}~\bibnamefont{Poplawska}},
  \bibinfo{author}{\bibfnamefont{R.}~\bibnamefont{Baranowski}},
  \bibnamefont{and} \bibinfo{author}{\bibfnamefont{T.}~\bibnamefont{Buchner}},
  \bibinfo{journal}{Chaos, Solitons and Fractals}
  \textbf{\bibinfo{volume}{11}}, \bibinfo{pages}{1061} (\bibinfo{year}{2000}).

\bibitem[{EPA({\natexlab{a}})}]{EPAPS03}
\eprint{See EPAPS Document No. E-PLEEE8-68-116309 for additional information.
  This document may be retrieved via the EPAPS homepage
  (\url{http://www.aip.org/pubservs/epaps.html}) or from \url{ftp.aip.org} in
  the directory /epaps/. See the EPAPS homepage for more information.}

\bibitem[{\citenamefont{Laguna et~al.}(1997)\citenamefont{Laguna, Mark,
  Goldberger, and Moody}}]{LAG97}
\bibinfo{author}{\bibfnamefont{P.}~\bibnamefont{Laguna}},
  \bibinfo{author}{\bibfnamefont{R.~G.} \bibnamefont{Mark}},
  \bibinfo{author}{\bibfnamefont{A.}~\bibnamefont{Goldberger}},
  \bibnamefont{and} \bibinfo{author}{\bibfnamefont{G.~B.} \bibnamefont{Moody}},
  in \emph{\bibinfo{booktitle}{Computers in Cardiology}}
  (\bibinfo{publisher}{IEEE Computer Society Press},
  \bibinfo{address}{Piscataway, NJ}, \bibinfo{year}{1997}),
  vol.~\bibinfo{volume}{24}, p. \bibinfo{pages}{673}.

\bibitem[{\citenamefont{Motulsky}(1995)}]{mot95}
\bibinfo{author}{\bibfnamefont{H.}~\bibnamefont{Motulsky}},
  \emph{\bibinfo{title}{Intuitive Biostatistics}} (\bibinfo{publisher}{Oxford
  University Press}, \bibinfo{address}{New York}, \bibinfo{year}{1995}).

\bibitem[{\citenamefont{Goldberger et~al.}(2002)\citenamefont{Goldberger,
  Amaral, Hausdorff, Ivanov, Peng, and Stanley}}]{PNAS02}
\bibinfo{author}{\bibfnamefont{A.~L.} \bibnamefont{Goldberger}},
  \bibinfo{author}{\bibfnamefont{L.~A.~N.} \bibnamefont{Amaral}},
  \bibinfo{author}{\bibfnamefont{J.~M.} \bibnamefont{Hausdorff}},
  \bibinfo{author}{\bibfnamefont{P.~C.} \bibnamefont{Ivanov}},
  \bibinfo{author}{\bibfnamefont{C.-K.} \bibnamefont{Peng}}, \bibnamefont{and}
  \bibinfo{author}{\bibfnamefont{H.~E.} \bibnamefont{Stanley}},
  \bibinfo{journal}{Proc. Natl. Acad. Sci. USA} \textbf{\bibinfo{volume}{99}},
  \bibinfo{pages}{2467} (\bibinfo{year}{2002}).

\bibitem[{\citenamefont{Jan\'{e} et~al.}(1997)\citenamefont{Jan\'{e}, Blasi,
  J.~Garcia, and Laguna}}]{jan97}
\bibinfo{author}{\bibfnamefont{R.}~\bibnamefont{Jan\'{e}}},
  \bibinfo{author}{\bibfnamefont{A.}~\bibnamefont{Blasi}},
  \bibinfo{author}{\bibfnamefont{J.}~\bibnamefont{J.~Garcia}},
  \bibnamefont{and} \bibinfo{author}{\bibfnamefont{P.}~\bibnamefont{Laguna}},
  in \emph{\bibinfo{booktitle}{Computers in Cardiology}}
  (\bibinfo{publisher}{IEEE Computer Society Press},
  \bibinfo{address}{Piscataway, NJ}, \bibinfo{year}{1997}),
  vol.~\bibinfo{volume}{24}, p. \bibinfo{pages}{295}.

\bibitem[{\citenamefont{Laguna et~al.}(1990)\citenamefont{Laguna, Thakor,
  Caminal, Jan\'{e}, and Yoon}}]{lag90}
\bibinfo{author}{\bibfnamefont{P.}~\bibnamefont{Laguna}},
  \bibinfo{author}{\bibfnamefont{N.~V.} \bibnamefont{Thakor}},
  \bibinfo{author}{\bibfnamefont{P.}~\bibnamefont{Caminal}},
  \bibinfo{author}{\bibfnamefont{R.}~\bibnamefont{Jan\'{e}}}, \bibnamefont{and}
  \bibinfo{author}{\bibfnamefont{H.~R.} \bibnamefont{Yoon}},
  \bibinfo{journal}{Medical \& Biological Engineering \& Computing}
  \textbf{\bibinfo{volume}{28}}, \bibinfo{pages}{67} (\bibinfo{year}{1990}).

\bibitem[{\citenamefont{Laguna et~al.}(1994)\citenamefont{Laguna, Jan\'{e}, and
  Caminal}}]{lag94}
\bibinfo{author}{\bibfnamefont{P.}~\bibnamefont{Laguna}},
  \bibinfo{author}{\bibfnamefont{R.}~\bibnamefont{Jan\'{e}}}, \bibnamefont{and}
  \bibinfo{author}{\bibfnamefont{P.}~\bibnamefont{Caminal}},
  \bibinfo{journal}{Computers and Biomedical Research}
  \textbf{\bibinfo{volume}{27}}, \bibinfo{pages}{45} (\bibinfo{year}{1994}).

\bibitem[{\citenamefont{Ivanov et~al.}(2001)\citenamefont{Ivanov, Amaral,
  Goldberger, Havlin, Rosenblum, Stanley, and Struzik}}]{iva01}
\bibinfo{author}{\bibfnamefont{P.~C.} \bibnamefont{Ivanov}},
  \bibinfo{author}{\bibfnamefont{L.~A.~N.} \bibnamefont{Amaral}},
  \bibinfo{author}{\bibfnamefont{A.~L.} \bibnamefont{Goldberger}},
  \bibinfo{author}{\bibfnamefont{S.}~\bibnamefont{Havlin}},
  \bibinfo{author}{\bibfnamefont{M.~G.} \bibnamefont{Rosenblum}},
  \bibinfo{author}{\bibfnamefont{H.~E.} \bibnamefont{Stanley}},
  \bibnamefont{and} \bibinfo{author}{\bibfnamefont{Z.~R.}
  \bibnamefont{Struzik}}, \bibinfo{journal}{Chaos}
  \textbf{\bibinfo{volume}{11}}, \bibinfo{pages}{641} (\bibinfo{year}{2001}).

\bibitem[{\citenamefont{McSharry et~al.}(2003)\citenamefont{McSharry, Clifford,
  Tarassenko, and Smith}}]{sha03}
\bibinfo{author}{\bibfnamefont{P.~E.} \bibnamefont{McSharry}},
  \bibinfo{author}{\bibfnamefont{G.~D.} \bibnamefont{Clifford}},
  \bibinfo{author}{\bibfnamefont{L.}~\bibnamefont{Tarassenko}},
  \bibnamefont{and} \bibinfo{author}{\bibfnamefont{L.~A.} \bibnamefont{Smith}},
  \bibinfo{journal}{IEEE Trans. Biomed. Eng.} \textbf{\bibinfo{volume}{550}},
  \bibinfo{pages}{289} (\bibinfo{year}{2003}).

\bibitem[{\citenamefont{Peng et~al.}(1995)\citenamefont{Peng, Havlin, and
  Stanley}}]{pen95}
\bibinfo{author}{\bibfnamefont{C.-K.} \bibnamefont{Peng}},
  \bibinfo{author}{\bibfnamefont{S.}~\bibnamefont{Havlin}}, \bibnamefont{and}
  \bibinfo{author}{\bibfnamefont{H.~E.} \bibnamefont{Stanley}},
  \bibinfo{journal}{Chaos} \textbf{\bibinfo{volume}{5}}, \bibinfo{pages}{82}
  (\bibinfo{year}{1995}).

\bibitem[{\citenamefont{Khan}(2002)}]{p16}
\bibinfo{author}{\bibfnamefont{I.~A.} \bibnamefont{Khan}},
  \bibinfo{journal}{Am. Heart J.} \textbf{\bibinfo{volume}{143}},
  \bibinfo{pages}{7} (\bibinfo{year}{2002}).

\bibitem[{\citenamefont{Abramowitz and Stegun}(1970)}]{abr70}
\bibinfo{author}{\bibfnamefont{M.}~\bibnamefont{Abramowitz}} \bibnamefont{and}
  \bibinfo{author}{\bibfnamefont{I.~A.} \bibnamefont{Stegun}},
  \emph{\bibinfo{title}{Handbook of Mathematical Functions}}
  (\bibinfo{publisher}{Dover}, \bibinfo{address}{New York},
  \bibinfo{year}{1970}).

\bibitem[{\citenamefont{Feller}(1971)}]{fel71}
\bibinfo{author}{\bibfnamefont{W.}~\bibnamefont{Feller}},
  \emph{\bibinfo{title}{An Introduction to Probability Theory and Its
  Applications, Vol. II}} (\bibinfo{publisher}{Wiley}, \bibinfo{address}{New
  York}, \bibinfo{year}{1971}).

\bibitem[{\citenamefont{Gradsteyn and Ryzhik}(1980)}]{gra80}
\bibinfo{author}{\bibfnamefont{I.~S.} \bibnamefont{Gradsteyn}}
  \bibnamefont{and} \bibinfo{author}{\bibfnamefont{I.~M.}
  \bibnamefont{Ryzhik}}, \emph{\bibinfo{title}{Table of Integrals, Series and
  Products}} (\bibinfo{publisher}{Academic Press}, \bibinfo{address}{San
  Diego}, \bibinfo{year}{1980}).

\bibitem[{EPA({\natexlab{b}})}]{EPAPS}
\eprint{See EPAPS Document No. E-PLEEE8-69-107405 for additional information. A direct link 
to this document may be mfound in the online article's HTML reference section. The document
may also be readed via via the EPAPS homepage
  (\url{http://www.aip.org/pubservs/epaps.html}) or from \url{ftp.aip.org} in
  the directory /epaps/. See the EPAPS homepage for more information. }

\end{thebibliography}

\end{document}